\documentclass[fleqn,twoside]{article}
\usepackage[]{espcrc2}

\readRCS
$Id: espcrc2.tex,v 1.2 2004/02/24 11:22:11 spepping Exp $
\usepackage{graphicx}
\usepackage[figuresright]{rotating}

\newcommand{\AmS}{{\protect\the\textfont2
  A\kern-.1667em\lower.5ex\hbox{M}\kern-.125emS}}

\def\bar{\overline}

\def\non{\nonumber}
\def\eqn{\begin{equation}}
\def\eqne{\end{equation}}
\def\eqa{\begin{eqnarray}}
\def\eqae{\end{eqnarray}}
\def\ary{\begin{array}}
\def\arye{\end{array}}
\def\dsc{\begin{description}}
\def\dsce{\end{description}}
\def\itm{\begin{itemize}}
\def\itme{\end{itemize}}
\def\enu{\begin{enumerate}}
\def\enue{\end{enumerate}}
\def\ct{\begin{center}}
\def\cte{\end{center}}
\def\GeV{{\rm GeV}}

\title{Multi-leg calculations with the {\tt GRACE/1-LOOP} system\\
--- Toward Radiative Corrections to
$e^+e^- \rightarrow \mu^-\bar{\nu} u \bar{d}$ ---}

\author{F. Boudjema\address{LAPTH, B.P.110, Annecy-le-Vieux F-74941,
                                France},
        J. Fujimoto\address{KEK, Oho 1-1, Tsukuba, Ibaraki 305-0801, Japan},
        T. Ishikawa$^b$
        T. Kaneko$^b$,
        K. Kato\address{Kogakuin University, Nishi-Shinjuku 1-24,
        Shinjuku, Tokyo 163-8677, Japan},
        Y. Kurihara$^b$,
        and
        Y.Shimizu$^b$}

\begin{document}

\begin{abstract}
We performed the calculation of the full $O(\alpha)$ corrections
to $ e^+e^- \rightarrow \mu^-\bar{\nu}u\bar{d}$ with the help of
the {\tt GRACE/1-LOOP} system. We discuss how a finite decay width
introduces a serious gauge invariance breaking, particularly for
infrared 5-point functions. This is related to the way the
reduction of those functions is performed and to the treatment of
the width in the reduction.

\vspace{1pc}
\end{abstract}

\maketitle

\vspace*{-9.9cm}
\noindent
KEK--CP--154 \\
LAPTH--1053
\vspace*{8.7cm}

\section{Introduction}
The first calculation of the full $O(\alpha)$ corrections to
$e^+e^- \rightarrow \nu \bar{\nu}H$ was presented at the
RADCOR/L\&L 2002 workshop\cite{ll2002}. Since then, several
authors published the radiative corrections to several important
$2 \rightarrow 3$ processes: $e^+e^- \rightarrow \nu
\bar{\nu}H$\cite{nnh1,nnh2,nnh3,nnh4}, $e^+e^- \rightarrow t
\bar{t}H$\cite{tth1,tth2,tth3,tth4}, $e^+e^- \rightarrow
ZHH$\cite{zhh1,zhh2}, $\gamma \gamma \rightarrow t
\bar{t}H$\cite{aatth}, $e^+e^- \rightarrow e^+e^-H$\cite{eeh}, and
$e^+e^- \rightarrow \nu \bar{\nu}\gamma$\cite{eeh}. Now, full EW
1-loop calculations are well under control for $2 \rightarrow 3$
processes in the SM. In this paper, we discuss the case for a $2
\rightarrow 4$ process.

As a first trial we take a typical LEP-2 process $ e^+e^-
\rightarrow \mu^-\bar{\nu}u\bar{d}$. Although a status report on a
similar attempt has appeared\cite{vicini}, there still is no
complete result.

\section{Motivation}

For LEP2 experiments, the Double Pole
Approximation(DPA)\cite{lep2} and  the fermion loop
scheme\cite{lep2} were used to predict the cross sections for
$e^+e^-
\rightarrow $ 4-fermions. They have the following features:(1)
Gauge invariance is guaranteed. (2) It was sensible to split all
10 tree diagrams(CC10) into the doubly resonant ones(CC03) and
others. The constant width is introduced in a naive way. In the
energy region at the future linear-collider, however, non-CC03
diagrams are not negligible. For example, at $\sqrt{s}=500$ GeV,
the tree level cross sections for CC03 and CC10 are 213.56 fb and
222.39 fb, respectively. The difference between them reaches 4\%.
Therefore, the size of the radiative corrections to CC10 should be
carefully estimated at the TeV energy region.

\section{Structure of the calculations}
{\tt GRACE/1-LOOP}\cite{grace-loop} has been used for the
calculation  which proceeds through the following steps: 1)
Evaluation of the numerators, for which the symbolic manipulation
system is used. In order to shorten the size of the matrix
elements, the fermion masses $m_e,m_{\mu}$, $m_{u},m_{d}$ are
neglected when integrating  over  phase space. 2) Evaluation of
the loop integrals. After 5- and 6-point integrals are reduced to
4-point functions, {\tt FF}\cite{ff} and other analytic formulas
are invoked for 4-, 3-and 2-point integrals. All masses are kept
in the loop integrals. 3) Construction of kinematics. Here masses are
also kept exactly.

\section{The reduction algorithm of $N(\geq 5)$ point functions}

\begin{figure}[htb]
    \begin{center}
    \includegraphics[width=6.5cm,clip=true]{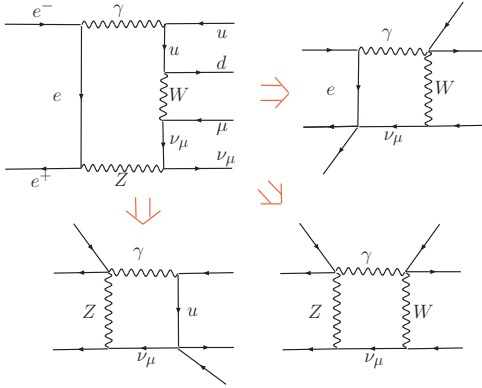}
    \end{center}
 \vspace*{-1.5cm}
 \caption{{\em Example of unstable 4-point functions obtained in the reduction of a 6-point
 function}}
    \label{fig:box}
\end{figure}

The amplitude with 5-point function\cite{denner,nnh2,zhh1} is
\small
\eqa
T^{(5)}&\propto&\int \frac{d^nl}{(2\pi)^ni}\frac{l_{\mu}l_{\nu}
\cdots l_{\rho}}{D_0 D_1 \cdots D_4},
\eqae
\normalsize
where $D_0=l^2-m_0^2$ and $D_i=(l+s_i)^2-m_i^2,i=1,\cdots,4$ and
$s_i$ is a combination of external momenta.

Multiplying the identities
\small
\eqa
g^{\mu\nu}=\sum_{i,j=1}^4 s_i^{\mu}(A^{-1})_{ij}s_j^{\nu},
\nonumber
\eqae
with the loop momentum $l$,
\eqa
l^{\mu}&=&\sum_{i,j}s_i^{\mu}(A^{-1})_{ij}(l \cdot s_j),
\nonumber\\
  &=&\frac{1}{2}\sum_{i,j}(A^{-1})_{ij}(D_j-D_0-\Delta_j)s_i^{\mu},
\eqae
\normalsize
where $A_{ij}=s_i\cdot s_j$ and $\Delta_i=s_i^2-m_i^2$,
we can reduce the numerator to
\small
\eqa
l_{\mu}l_{\nu} \cdots l_{\rho}=\frac{1}{2}\sum_{i,j}(A^{-1})_{ij}
(D_i-D_0-\Delta_i)s_j^{\mu}l_{\nu} \cdots l_{\rho}.
\label{eq:id1}
\eqae

\normalsize
This method is also applicable for the reduction of a 6-point
function to a sum of 4-point functions. There is also a  standard
reduction method where the following identity is used;
\eqa
  l^2=\sum_{i,j}(l \cdot s_i)A_{ij}(l \cdot s_j).
\label{eq:id2}
\eqae
Because this method not only raises the rank of the momentum
tensor in the numerator but makes the resultant source codes very
lengthy compared with Eq.\ref{eq:id1}, {\tt GRACE/1-LOOP} uses
this identity only as an option.  As mentioned the system employs
the {\tt FF} package for the evaluation of 4-point functions. It
happens, however, that this  package leads to numerical
instabilities or inconsistency in some cases having to do with
some infra-red boxes. This occurs for instance when an internal
massless particle is involved in some non-IR boxes like those
obtained from a 6-point function as shown in Fig.1. Some infra-red
boxes also need to be regulated by the introduction of a width for
the particle($W$ or $Z$) circulating in the loop. We implement a
constant fixed width for such cases. For all these particular
cases we use special in-house routines. For the computations we
performed till now these routines have worked quite well for any
$2\rightarrow2$ and all $2\rightarrow 3$ one-loop processes.
However, for the problem at hand, we have met a difficulty in
carrying out the evaluation of a $5$-point function when the $W$
width is required.

\small
\begin{center}
\begin{table*}[htb]
\caption{{\em Stability of the matrix element squared, Eq.~5, against the changes
of $C_{UV}$ and $\lambda$.}}
\begin{tabular}{|c|c|c|c|} \hline
graph&$C_{UV}$&$\lambda=10^{-18}$ GeV & $\lambda=10^{-21}$ GeV\\
\hline
full      &0
& {\small -0.85388129300841993023002220162E-03}
& {\small -0.85388129300794622839025322460E-03} \\
full      &100
& {\small -0.85388129300841993023002220166E-03}
& {\small -0.85388129300794622839025322465E-03} \\
prod.&0
& {\small -0.85388086448063959950419248157E-03}
& {\small -0.85388086990548652947683563335E-03} \\
prod.&100
& {\small -0.85388086452253511266116024563E-03}
&{\small  -0.85388086994738204263380339741E-03} \\
\hline
\end{tabular}
\end{table*}
\end{center}

\normalsize

\section{1-loop diagrams for $e^+e^- \rightarrow
\mu^-  \bar{\nu}u\bar{d}$}

We introduce two sets of diagrams, the {\itshape full set\/}
and the {\itshape production set\/}.
In the standard model and within the non-linear gauge fixing
conditions\cite{boudjema},
one counts 44 tree-level and 6094 1-loop diagrams.
This defines the {\itshape full set\/}.

When the  masses of  the electron, $\mu$, $u$- and $d$-quark are
ignored in the numerator of the matrix elements, we obtain the
{\itshape production set\/} which consists of 10 tree
diagrams(CC10) and 668 1-loop diagrams (with 88 pentagon and 60
hexagon graphs).

\section{Check of codes}

The following set of input parameters are used in the calculation:
Throughout this paper, the results are expressed in terms of the fine
structure constant in the Thomson limit
$\alpha_{QED}^{-1}=137.0359895$
and the $Z$ mass $M_Z=91.1876$GeV. The on-shell renormalization
scheme uses $M_W$ as an input parameter. Nevertheless the numerical
value of $M_W$ is derived through $\Delta r$\cite{hioki}.
$M_W$ thus changes as a
function of $M_H$. We take $m_u=m_d=63$ MeV, $m_s=92$ MeV,
$m_c=1.5$ GeV, $m_b$=4.7GeV. This set of masses gives a ``perturbative"
value of $\alpha(M_Z)$ compatible with the current value derived
experimentally. We also take $m_{top}$=180GeV. With these values, we
find $M_W$ =80.4163 and $\Gamma_Z$=2.4952GeV for $M_H$=120GeV.
$\Gamma_W$=2.118GeV is taken from PDG.

The results of the calculation are checked by performing four
kinds of tests at few points in phase space. For this we worked
with the {\itshape full set\/} of diagrams in quadruple precision.
The first check is the ultraviolet finiteness test. The regulator
constant $C_{UV}=1/\varepsilon -\gamma_E+\log 4\pi,
n=4-2\varepsilon$ is kept in the matrix elements. In varying this
parameter $C_{UV}$, we found that the result is stable with an
accuracy reaching $29$ digits. Infrared(IR) finiteness is also
checked by introducing a fictitious photon mass $\lambda$, treated
as an input parameter in the code. The sum of loop and
bremsstrahlung contributions stays stable against variations in
$\lambda$ with an accuracy of $12$ digits. Table 1 shows the
stability for both cases of {\itshape full set\/} and {\itshape
production set\/}(prod.). For the latter the accuracy worsens to
order $O(m_{\mu}^2/s)$ because the masses of the fermions are
neglected in the numerator.

The third check relates to the independence on the parameter $k_c$
which is a soft photon cut parameter that separates soft photon
radiation(analytical formula) and the hard photon performed  by
the Monte-Carlo integration. Gauge parameter independence of the
result is performed as the last check through a set of five gauge
fixing parameters. For the latter a generalized non-linear gauge
fixing condition\cite{boudjema} has been chosen. In total the
amplitude contains 5 free parameters:
$\tilde{\alpha},\tilde{\beta},\tilde{\delta},\tilde{\varepsilon}$
and $\tilde{\kappa}$. In this check we set $\Gamma_W=\Gamma_Z=0$
and keep all masses in the numerator. The whole matrix element
shows no dependence on any of the gauge parameters in more than 21
digits.

\small
\noindent
\begin{center}
\begin{table*}[htb]
\caption{{\em Cancellation between each 4-point function
in Eq.(8)  at a phase point with the matrix elements of Fig.2}}
\begin{tabular}{|c|c|c|}
\hline
box& $\Gamma_W=0$& $\Gamma_W\neq0$\\
&(real,~imag)&(real,~imag)\\
\hline
1   & (~$-17.69217505$, ~~~66.88369265) & ( ~~$-4.22744975$, ~~2.9465712570) \\
2   & ($-929.96695924$, ~1372.81890328) & ( $-120.36833308$, ~39.2204483424) \\
3   & (~782.70500402, -5690.37576950) & ( ~321.80994609, $-283.271079293$) \\
4   & (~410.83071602, ~4557.14502347) & ( $-199.63435655$, ~275.873356743) \\
5   & ($-245.31384970$, ~-301.36098220) & ( ~~~0.65534291, ~$-27.683645798$) \\
\hline
sum & (~~~0.56273604, ~~~~5.11086770) & ( ~~$-1.76485038$, ~~~7.085651250) \\
\hline
\end{tabular}
\end{table*}
\end{center}
\normalsize
\begin{figure}[bht]
    \begin{center}
    \includegraphics[width=8.5cm,clip=true]{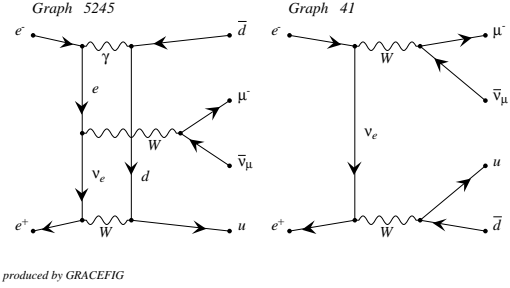}
 \vspace*{-1.7cm}
 \caption{{\em An IR 5-point function and a tree diagram.}}
    \end{center}
    \label{fig:box}
\end{figure}
 \vspace*{-1.6cm}
\section{Status of calculation}
All the calculations were done in quadruple precision. When we
sampled points in phase space  to carry out the  MC integration,
we encountered  an unstability  with IR 5-point diagrams. In Fig.2
we show such an example with a product of a 1-loop IR diagram and
a tree diagram. It is to be noted that the 1-loop graph is
obtained from the tree by adding a photon that joins $e^-$ and
$u$.
A 5-point diagram is written as a sum of {\it five} 4-point
integrals. For the product of the two diagrams (the labelling is
assigned automatically by the system), we then write
\eqa
&&[\hbox{IR-loop}]^{(5)} \equiv 2 T_{5245}T_{41}^{\dagger} \non\\
            &=& I^{(4)IR}_1+I^{(4)IR}_2+I^{(4)}_3+I^{(4)}_4+I^{(4)}_5,
    \label{eq:sig}
\eqae
which contains {\it two} infrared divergent boxes. Currently the
widths  $\Gamma_W$ and $\Gamma_Z$ are retained only in the IR part
of the box integrals in such a way
\begin{eqnarray}
I^{(4)IR}_1\sim{1\over D_W(q^2)}
\left[\log{\lambda\sqrt{\tilde M_W^2}\over D_W(q^2)}
\log(\cdots)+\cdots\right],
\end{eqnarray}
\eqa
D_W(q^2)&\equiv& q^2-\tilde M_W^2 =q^2-M_W^2+i\Gamma_WM_W \non
\eqae
and also in the reduction formulas for 4-points.
Without the
width, 3 digits cancellation is observed among the boxes, as shown
in Table~2 at some phase space point near the $W$-pole,
\eqa
&&(q^2-M_W^2+i\Gamma_WM_W)_{ud}\non\\
&&=-169.3611+167.1833i \qquad({\GeV}^2)
\eqae
On the other hand, when $\Gamma_W\neq0$, Table~2 shows less
cancellation. This led to a bad  convergence of the phase space
integration as we will  clearly demonstrate below.
Though the
mechanism for this cancellation with zero width is not fully
understood now, it is highly probable that an identity including
\begin{eqnarray*}
   \{
          (l.p_1) \varepsilon(p_2,p_3,p_4,p_5)
        - (l.p_2) \varepsilon(p_1,p_3,p_4,p_5)
  \\
     + (l.p_3) \varepsilon(p_1,p_2,p_4,p_5)
        - (l.p_4) \varepsilon(p_1,p_2,p_3,p_5)
  \\
     + (l.p_5) \varepsilon(p_1,p_2,p_3,p_4)
     \} \times
 (l.p_6)  = 0
\end{eqnarray*}
causes the trouble. This identity relates the five box integrals
among each other. No such phenomenon took place for 6-point
diagrams nor non-IR 5-points. Also we did not encounter such
difficulty in all $2\to3$ processes we studied till now. The
number of independent fermion lines may be related to this
undesirable situation. As mentioned above, {\tt GRACE/1-LOOP}
adopts Eq.\ref{eq:id1} for the reduction. We found, however, that
when we used Eq.\ref{eq:id2} no such problem happened.

\section{Test run of integration}
In order to confirm that an improper treatment of the width in the
infrared divergent 5-point function is the cause of bad
convergence, we temporarily took the following {\it ad hoc}
regularization in the loop amplitude
\eqa
      &&\sigma^{(5)}= 
        2\Re \left( T_{\hbox{IR-loop}^{(5)}}T_{\hbox{tree}}^{\dagger}\right)
                                                      \non\\
      &&\Longrightarrow
      2\Re\left[ {q^2-M_W^2\over q^2-\tilde M_W^2}
      \left(T_{\hbox{IR-loop}^{(5)}}\right)_{\Gamma_W=0} 
       T_{\hbox{tree}}^{\dagger} \right]
      \non\\
\label{eq:adhoc}
\eqae
and integrated all the diagrams in the phase space. The
integration is carried out by {\tt BASES}. Those diagrams which
contribute less than 0.01 fb were omitted to leave 361 1-loop
diagrams. Hard photon emission is also included.

At $\sqrt{s}=500~\GeV$, $\sigma_{tree}=222.39\pm 0.01$fb. Table~3
shows the results of the integrated cross section for each set of
$n$-point diagrams with 50,000 sampling points,
together with their MC error. The column  ``original(IR)" shows
the results of the diagrams related to the infrared divergence as
extracted normally through Eqs.~\ref{eq:sig}-6 or an equivalent
form for the other $n-$point IR diagrams. On the other hand ``{\it
ad hoc}(IR)" is the result after making the modification in
Eq.\ref{eq:adhoc} for the same sets of diagrams as
``original(IR)". The last column contains those diagrams which do
not have any infrared divergence. For the case of 5-point
functions of the type  ``original(IR)", the integrated error is
extremely huge due to the mismatch between its corresponding
4-point parts as mentioned earlier. After applying
Eq.\ref{eq:adhoc}, there is a drastic improvement in the error and
the central value of the integration is also shifted. Other
contributions from $n$-point functions remain the same within the
integration error if one applies the ``{\it ad hoc}" prescription.
Table~3 suggests that a proper treatment of the width in 5-point
functions is crucial to improve the situation but one still needs
to find  a proper  justification for the factorization.

\small

\vspace{5mm}
\noindent
Table 3

\noindent
{\em Integrated cross sections in {\tt fb} showing the
contribution of each set of $n$-point diagrams for the test run.
The parentheses show the MC integration error.}
\begin{center}
\begin{tabular}{|c|c|c|c|}
\hline
graph & original(IR) & {\it ad hoc}(IR) & non-IR\\
\hline
6-pnt& \texttt{~-921(~~16)} & \texttt{~-914(~6)} &  small   \\
5-pnt& \texttt{-4221(2224)} & \texttt{+2729(17)} & \texttt{-80(10)} \\
4-pnt& \texttt{-4041(~~26)} & \texttt{-3999(26)} & \texttt{+216(7)} \\
3-pnt& \texttt{~+735(~~~6)} & \texttt{~+735(~6)} & \texttt{-258(2)} \\
2-pnt&                      &                    & \texttt{-27(0.3)} \\
self & \texttt{~-104(~~~2)} & \texttt{~-104(~2)} & \texttt{-9(0.08)} \\
cnt  & \texttt{~+305(~~26)} & \texttt{~+305(26)} &  small   \\
\hline
soft & \multicolumn{2}{c|}{\texttt{~+990(~8)}} & \\
\hline
hard &  & & \texttt{+461(0.5)} \\
\hline
total& \texttt{-7257(2223)} & \texttt{~-258(~9)} & \texttt{+302(9)} \\
\hline
\end{tabular}
\end{center}
\normalsize

\section{Summary}
One-loop amplitudes of the full $6094$ diagrams for
 $e^+e^- \rightarrow \mu^- \bar{\nu} u\bar{d}$ were generated
by the {\tt GRACE/1-LOOP} system. Non-linear gauge invariance has
shown the consistency of the full set of amplitudes and the system
itself. A new reduction algorithm from a 6-point function to
4-point function works well. A finite decay width brings a serious
breaking of gauge invariance, particularly for 5-point infrared
integrals. It is clear now that the radiative corrections to
$2\to4$ processes are calculable, though more improvements are
inevitable.

\section{Acknowledgment}

This work is part of a collaboration between the Minami-Tateya
group and  LAPP/LAPTH. D. Perret-Gallix and G. B\'elanger deserve
special thanks for their contribution. The author(J.F.) also would
like to acknowledge the local organizing committee of LOOPS/LEGS
2004 for a stimulating workshop and for their nice organization.
This work was supported in part by Japan Society for Promotion of
Science under the Grant-in-Aid for scientific Research B(no.
14340081) and GDRI of the French National Centre for Scientific
Research (CNRS).

\end{document}